\documentclass[10pt]{article}

\usepackage{makeidx}
\usepackage{bbding}
\usepackage{dsfont}
\usepackage{amsmath}
\usepackage{graphicx}
\usepackage{booktabs} 
\usepackage{multirow}
\usepackage{subcaption}
\makeatletter
\newcommand{\printfnsymbol}[1]{%
  \textsuperscript{\@fnsymbol{#1}}%
}
\makeatother

\usepackage{authblk}

\begin{document}

\title{Towards a phylogenetic measure to quantify HIV incidence}

\author[1,2]{Pieter Libin \thanks{equal contribution}}
\author[5,6]{Nassim Versbraegen\printfnsymbol{1}}
\author[2,3]{Ana B. Abecasis}
\author[4]{Perpetua Gomes}
\author[1,5]{Tom Lenaerts}
\author[1]{Ann Now\'{e}}

\affil[1]{Artificial Intelligence Lab, Department of computer science, Vrije Universiteit Brussel, Brussels, Belgium}
\affil[2]{Department of Microbiology and Immunology, Rega Institute for Medical Research, KU Leuven - University of Leuven, Leuven, Belgium}
\affil[3]{Global Health and Tropical Medicine, GHTM, Instituto de Higiene e Medicina Tropical, IHMT, Universidade Nova de Lisboa, UNL, Lisboa, Portugal}
\affil[4]{Laboratorio Biologia Molecular, LMCBM, SPC, HEM, Centro Hospitalar Lisboa Ocidental}
\affil[5]{Machine Learning group, Université Libre de Bruxelles, Boulevard du Triomphe CP212, 1050 Bruxelles, Belgium}
\affil[6]{Interuniversity Institute for Bioinformatics in Brussels, ULB-VUB, 1050 Brussels, Belgium}
\date{}

\maketitle

\begin{abstract}
One of the cornerstones in combating the HIV pandemic is being able to assess the current state and evolution of local HIV epidemics. This remains a complex problem, as many HIV infected individuals remain unaware of their infection status, leading to parts of  HIV epidemics being undiagnosed and under-reported.
To that end, we firstly present a method to learn epidemiological parameters from phylogenetic trees, using approximate Bayesian computation (ABC). The epidemiological parameters learned as a result of applying ABC are subsequently used in epidemiological models that aim to simulate a specific epidemic.
Secondly, we continue by describing the development of a tree statistic, rooted in coalescent theory, which we use to relate epidemiological parameters to a phylogenetic tree, by using the simulated epidemics.  
We show that the presented tree statistic enables differentiation of epidemiological parameters, while only relying on phylogenetic trees, thus enabling  the construction of new methods to ascertain the epidemiological state of an HIV epidemic. 
By using genetic data to infer epidemic sizes, we expect to enhance understanding of the portions of the infected population in which diagnosis rates are low. 


\end{abstract}

\section{Introduction}

About 37 million people are currently infected with HIV and an estimated 35 million people have died due to the effects of AIDS (the eventual result of HIV infection) since the beginning of the epidemic at the start of the twentieth century \cite{UNAIDS_fact}.
Global efforts have ensued to enhance the collection, dissemination and accessibility of epidemiological data related to HIV. 
One of the most burdensome aspects in curtailing the spread of HIV emerges from infected individuals being unaware of their infection status. This stems from the fact that a host can be infected for many years before noticing any symptoms \cite{hightow2009identifying,burton2002follicular,theys2018impact}. As a result, a significant fraction of the HIV infected population remains undiagnosed, hampering effectiveness of interventions and assessment of further developments of the epidemic. Consequently, methods that deliver a well-founded estimate of the number of HIV infected individuals are paramount \cite{begier2010undiagnosed}. Such an estimate enables deduction of the number of undiagnosed infected individuals. 
State-of-the-art methods that aim to provide estimates of the size of HIV epidemics generally consist of applying compartment models to routine surveillance data to estimate the number of infected individuals (i.a. number of new diagnoses over time and CD4$^+$ cell counts) \cite{van2015estimating,mammone2016many}.

An abundance of clinical data is available in the context of HIV epidemics, as upon diagnosis a number of tests are performed and the results thereof collected. One of those tests comprises of assessing the specific genotype of the virus infecting a patient \cite{vajpayee2011current}. To that extent, the genetic sequence of the virus is determined. 
As a result, a vast amount of HIV sequences have been collected over the last decades. 

The main benefit of developing a method to quantify a HIV epidemic that relies on genetic data is to gain insight into the specific sub-populations that contain a high rate of undiagnosed individuals, thus allowing for more effective health policies, through diagnosis strategies that are directed towards these particular sub-populations.  \\
We validate our research on the HIV-1 epidemic in Portugal (see Section~\ref{sec:port_data}). We therefore first present inference of the epidemiological parameters of said epidemic by applying approximate Bayesian computation (ABC) \cite{sunnaaker2013approximate}. We apply ABC to fit a model that contains the epidemiological parameters in question (see Section~\ref{sec:simulcalib}). 
We further show that calibrating simulations to specific epidemics is essential, as the epidemiological dynamics has an important impact on the shape of the phylogenetic tree (see Section \ref{sec:results}). 
We then construct a tree statistic that enables differentiation of epidemiological parameters based on phylogenetic trees (see Section~\ref{sec:tzscore}) and evaluate it on a set of epidemiological simulations (see Section~\ref{sec:tzscore_results}).


\section{Background}


\subsection{Phylogenetic trees}

Phylogenetic trees represent evolutionary relationships between organisms. 
A rooted phylogenetic tree consists of a root, internal nodes, leaves and branches interconnecting nodes with other nodes and leaves.
The branches of a phylogenetic tree indicate a measure of distance between the organisms represented by their respective leaves. This distance can be based on the amount of genetic change or can represent natural time, by using a molecular clock \cite{mark2009evolution}. A smaller path between two nodes (i.e. traversing the tree through the nodes from one leaf to another) thus suggests a stronger evolutionary relatedness \cite{salemi2009phylogenetic}.\\

\subsection{Epidemiological models}

Compartment models are one of the most popular concepts stemming from mathematical epidemiology \cite{brauer2008compartmental}, 
Compartment models aim to capture population dynamics by stratifying individuals into different compartments, through some difference in state. They include transitions between compartments over time, representing changes in state in the modelled population. 
We illustrate this with the SIR Model \cite{kermack1927contribution}, which consists of three coupled non-linear ordinary differential equations, representing the change in each compartment over time; 
\begin{equation}
    \dot{S} = -{\beta}SI
\end{equation}
\begin{equation}
   \dot{I} = {\beta}SI -{\gamma}I
\end{equation}
\begin{equation}
    \dot{R} = {\gamma}I
\end{equation}
With $\beta$ the infection rate, $\gamma$ the recovery rate and $t$ time. 

\begin{figure}
\centering
\includegraphics[width=0.6\linewidth]{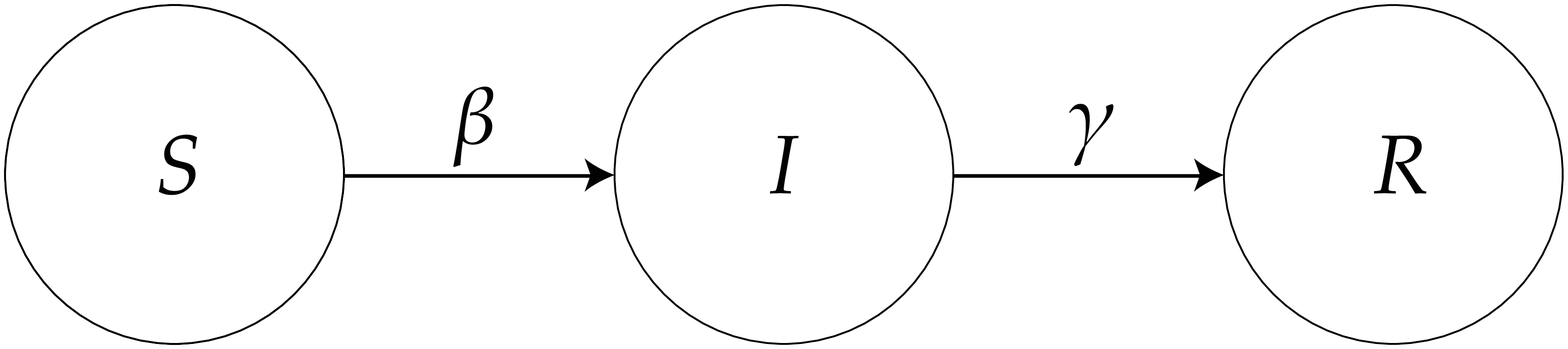}
\caption{SIR compartment model, compartments Susceptible ($S$), Infected ($I$) and Recovered ($R$) represented by circles, transitions between them by arrows, denoted by their associated rates ($\beta$ and $\gamma$)}
\label{fig:SIR}
\end{figure}

The model consists of 3 compartments; Susceptible ($S$), Infected ($I$) and Recovered ($R$) (or removed) and transitions between those compartments (see Figure \ref{fig:SIR}). $R$, $S$ and $I$ represent the number of people in the respective compartment at a certain point in time. The rates $\beta$ and $\gamma$ depend on different factors (i.a. population density, infectiousness and pathology). 

The model and extended versions thereof are especially relevant when trying to answer health policy questions. 
In our work, such models will be used to simulate specific HIV epidemics.

\subsection{Detecting missed infections in phylogenetic trees.}
The starting point for our tree statistic was the method presented in  \cite{Stacy2016}, which uses the coalescent \cite{kingman1982coalescent} to provide an indication of the extent to which samples are missing, or overly present in a specific phylogenetic tree.

Equation \ref{eq:full_unsampled} specifies how to calculate node probabilities for each node $j$ in a tree \cite{Stacy2016}.  

\begin{equation}\label{eq:full_unsampled}
\pi^j(t,k,N) = \sum^{j}_{i=1}k^{s}_{i} Pr_i(C_{lineage}) 
\end{equation}
With $i$ the interval index, $t_i$ the interval length, $k$ the number of lineages to coalesce in an interval, $N$  the population size and $Pr_i(C_{lineage})$ the lineage coalescence probability.\\

\begin{equation}\label{eq:z_score_unsampled}
Z^j \; \propto \;  n\pi^j - np^j
\end{equation}

To be able to gauge to what extent the population making up the tree under investigation is under- or oversampled, one can convert the node probabilities to z-scores, by using equation \ref{eq:z_score_unsampled}. The equation quantifies the relation between number of leaves expected at a node and number of observed leaves for that node \cite{Stacy2016}.


\subsection{Portuguese HIV-1 epidemic}
The first reports of HIV-1 diagnoses in Portugal go back to 1983. By 2014, 53072 diagnoses had been reported \cite{diniz2015portugal}. As in most European HIV epidemics, subtype B is the most prevalent HIV-1 subtype, followed by subtype G which is atypical in a European setting \cite{abecasis2013}. Within the "men who have sex with men" risk group, the number of yearly diagnoses show a mild but steady increase, while modes of transmission have transformed from intravenous drug use being the main cause of new infections to infections stemming from heterosexual sex in the period 2000-2014 \cite{diniz2015portugal}. This change is associated with lower diagnoses rates overall (consistent with diagnoses in Europe in general), possibly illustrating the beneficial results of a health policy implementation that was particularly effective for intravenous drug-users (e.g. through providing single-use needles). We apply our research on data stemming from this epidemic.

\section{Methods and Materials}

\subsection{Portuguese Data}
\label{sec:port_data}

The data used in the experiments was made available through a HIV-1 resistance database from \textit{Hospital Egas Moniz}. 
Henceforth, we will refer to the used data as 'Portuguese data' for the sake of brevity. \\
Data storage and querying was achieved through the RegaDB system \cite{libin2013regadb}. Said system allows for complex querying, which was key to assemble all the relevant data in an efficient manner. 
We proceeded by querying genetic sequences of HIV-1 belonging to distinct patients. 
In doing so, we assembled three genetic sequence sets, differing in the HIV-1 subtypes they contain. The first set only holds subtype B sequences (n= 2216) . The second set is made up of subtype G sequences (n=1961). And a final set (n= 6079) that does not take into account the specific subtype (and thus also includes other subtypes besides B and G). HIV-1 sequences were classified using the Rega typing tool \cite{pineda2013automated,alcantara2009standardized}. Each patient only contributed one genetic sequence to a set, if multiple sequences were associated with a single patient, the sequence established first was selected. \\
For each set, an alignment was created using MAFFT \cite{mafft_align}. The resulting alignments were then used to infer phylogenetic trees using maximum likelihood trough RAxML \cite{stamatakis2014raxml}. 
RAxML was used with the GTR-$\gamma$ model. A maximum likelihood tree was constructed and subsequently annotated through bootstraps. Bootstrapping was halted automatically based on extended majority rule consensus trees (i.e. autoMRE). 
In order to preserve the confidential nature of the employed patient data, tree inference was carried out on local computers exclusively and only anonymized patient data was used.

\subsection{Simulation calibration \label{sec:simulcalib}}

In order to validate our tree statistic on a real world epidemic, being able to generate simulation data that was plausible with regard to the real world epidemic was essential. We thus proceeded by inferring relevant epidemiological parameters in order to calibrate subsequent simulations. 
To that end, we opted to use ABC \cite{sunnaaker2013approximate} to learn said parameters, this approach was inspired by the work presented in \cite{poon2015phylodynamic}.
ABC is closely related to Markov chain Monte Carlo (MCMC), but unlike MCMC, does not require the calculation of exact likelihoods, which can be intractable for complex models \cite{poon2015phylodynamic,libin2018bayesian}.
Learning the relevant epidemiological parameters in an ABC setting requires the presence of some distance measure, as an alternative to the exact likelihoods used in MCMC approaches. Taking into account that the available epidemic data we want to infer parameters from exists in the form of a phylogenetic tree, and the possibility of generating new phylogenetic trees through simulation, we employ a kernel method developed by Poon \cite{poon2013kernel} as a distance measure between two trees. In concreto, we rely on a specified compartment model (see Figure \ref{fig:colSIR} for the used model) that enables the generation of trees.

By using the aforementioned kernel method, we assume that correspondence in trees reflect similarities between the model and the epidemic underlying the observed tree. In order to explore the possible parameters of the specified model efficiently, ABC is used. In essence, ABC varies parameter values in order to simulate more data by using the proposed parameter values in a specified model and aims to minimise the distance between the newly generated data and the observed data (in this case using the kernel method as a distance measure) \cite{sunnaaker2013approximate,poon2015phylodynamic}.
Table \ref{table:ABCpars} shows the parameters used in our ABC application.
\begin{table}
\centering

\begin{tabular}{c | c | c | c | c}
\toprule
\multirow{1}{*}{Parameter} & \multirow{1}{*}{Range} & \multirow{1}{*}{$\sigma$} & \multirow{1}{*}{Initial} & \multirow{1}{*}{Prior}\\
\midrule
$N$ & [$10^3$ - $10^6$] & $10^3$ & $10^4$ & $X=e^{\mu +\sigma Z}$, $\mu=0.5$ and $\sigma=10000$ \\

$\beta_i$ & [$10^{-3}$ - $10$] & $0.05$ & $0.5$& $X=e^{\mu +\sigma Z}$, $\mu=1.0$ and $\sigma=0.01$\\

$\gamma_i$ & [$0$ - $5$] & $0.01$& $0.1$ & $X=e^{\mu +\sigma Z}$, $\mu=1.0$ and $\sigma=0.01$\\

$\mu$ & [$0$ - $1$] & $0.002$ & $0.02$ & $X=e^{\mu +\sigma Z}$, $\mu=1.0$ and $\sigma=0.01$\\

\bottomrule
\end{tabular}
\caption{Parameters used in ABC kernel method SI model, $X=e^{\mu +\sigma Z}$ represents the log-normal distribution, $\beta_i$ being the infection rate, $\gamma_i$ the mortality rate, $\mu$ the mortality rate from natural causes, $N$ the population size.}
\label{table:ABCpars}
\end{table}

\subsection{Simulations}
\begin{figure}
\centering
\includegraphics[width=0.4\linewidth]{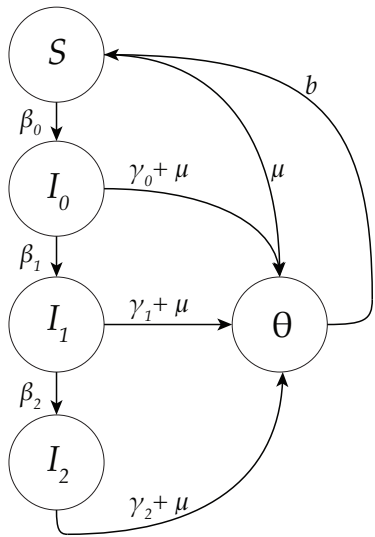}
\caption{extended SIR compartment model, compartments Susceptible ($S$), three Infected stages ($I_i$) and Deceased ($\theta$) represented by circles, transitions between them by arrows, denoted by their associated rates (infections $\beta_i$, death through infection $\gamma_i$, death trough natural causes $\mu$ and births $b$)}
\label{fig:colSIR}
\end{figure}

rcolgem \cite{rasmussen2014phylodynamicColgem,volz2012complexColgem} was used to simulate epidemics, based on parameter ranges obtained from application of the ABC-kernel method. 
Each simulation set consisted of 1000 simulations, outputting phylogenetic trees, and a log of the population dynamics over time. The used model (based on \cite{Volz2014ColgemHivRates,Volz2015ColgemHiv}) is given by the following equations, and is illustrated in Fig. \ref{fig:colSIR}.
\begin{equation*}
\dot{S} = bN - \mu S- (\beta_0I_0+\beta_1I_1+\beta_2I_2)\frac{S}{N}
\end{equation*}
\begin{equation*}
\dot{I_0} =(\beta_0I_0+\beta_1I_1+\beta_2I_2)\frac{S}{N} - (\mu + \gamma_0)I_0
\end{equation*}
\begin{equation*}
\dot{I_1} = \gamma_0I_0 - (\mu + \gamma_1)I_1
\end{equation*}
\begin{equation*}
\dot{I_2} = \gamma_1I_1 - (\mu + \gamma_2)I_2
\end{equation*}

The model is an extension of a SIR model, and consists of three infection stages and includes births and deaths (i.e. conceptual addition and removal of simulated individuals) without a recovery state.
Used parameter values, determined from the results of ABC application, are as follows; $\gamma_0 = 0.045$, $\gamma_1 = 0.14$, $\gamma_2 = 0.5$ $\mu = 0.001$, $\beta_0 = 0.12$, $\beta_1 = 0.03$ and $\beta_2 = 0.009$ while $S_0$ is varied between $145000$ and  $157000$ and the sample size (i.e. the number of leaves in the tree) between $1000$ and $12000$.  Parameters were sampled between the specified ranges using Latin hypercube sampling. 
The goal being to find parameters for the simulation engine that result in phylogenetic trees that are similar to the ones inferred from the Portugal data, in order to match the underlying epidemic.

\subsection{Constructing the $T_z$-score}
\label{sec:tzscore}
The starting point in constructing our tree statistic was the method described in  \cite{Stacy2016}. 
The result of applying said method on a tree is an annotated tree, which includes node probabilities for each node in the tree. In order to construct our tree statistic from such an annotated tree, we devised a procedure to infer information about the population on the basis of z-scores.
We proceeded by defining a statistic that demonstrates the overall extent to which a tree is over- or undersampled. We call this statistic $T_z$. 
To convert the obtained annotated tree to a single statistic value, we rely on the z-score in the root of the tree, as node probabilities are by definition propagated to the root node;  $T_z=\frac{Z^r}{s}$, with $Z^r$ the z-score of the root and $s$ the number of samples (i.e. leaves) making up the tree. In order to apply the method, a tree and a $N$ need to be specified. Through experimental analyses, we found that a large $N$ is necessary to obtain informative results, we thus specified $N=10^5$.


\section{Results}
\label{sec:results}

\subsection{Approximate Bayesian computation}
\begin{figure}
\centering
\includegraphics[width=0.7\linewidth]{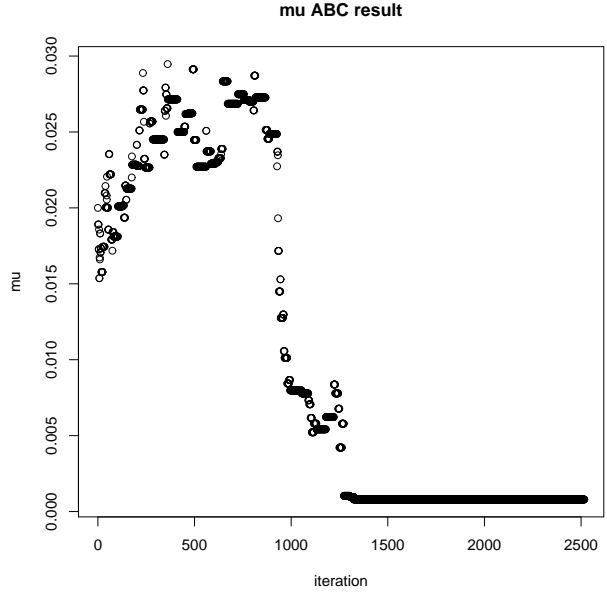}
\caption{ABC kernel method result, $\mu$ over different iterations} 
\label{fig:muABC}
\end{figure}

Figure \ref{fig:muABC} presents the results of ABC application for the $\mu$ parameter. The figure indicates that the ABC chain converged after about 1500 iterations and had thus learned plausible values for said parameters with regard to the specified model.

\subsection{Phylogenetic tree assessment}
We first present a visual comparison of tree topologies between the tree inferred from the Portuguese dataset, a tree obtained through an ABC application calibrated simulation and a PANGEA \cite{Ratmann2017PANGEA} simulation, which aims to model the HIV epidemic in sub-Saharan Africa
. If topologies show major discrepancies, we assume this indicated the simulations are not well calibrated with regard to the actual epidemic, while a relatively corresponding topology would indicate simulations resembling the actual epidemic. The PANGEA tree serves as an example of topology difference when comparing different epidemics.  

\begin{figure}
\begin{subfigure}{.45\textwidth}
\centering
\includegraphics[width=0.7\linewidth]{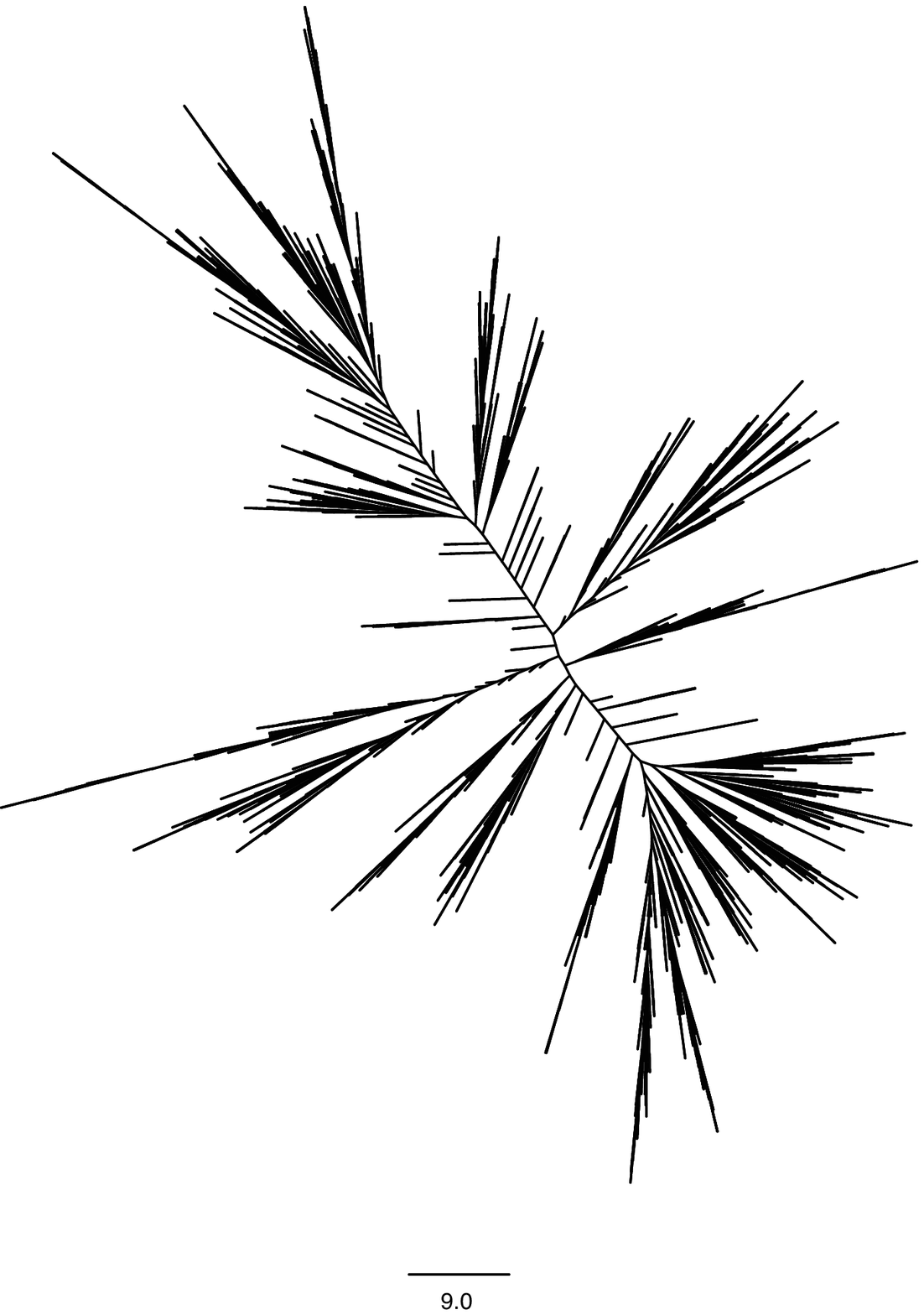}
\caption{Cladogram of tree inferred from Portuguese dataset}
\label{fig:port_tre}
\end{subfigure}\hfill
\begin{subfigure}{.45\textwidth}
\centering
\includegraphics[width=\linewidth]{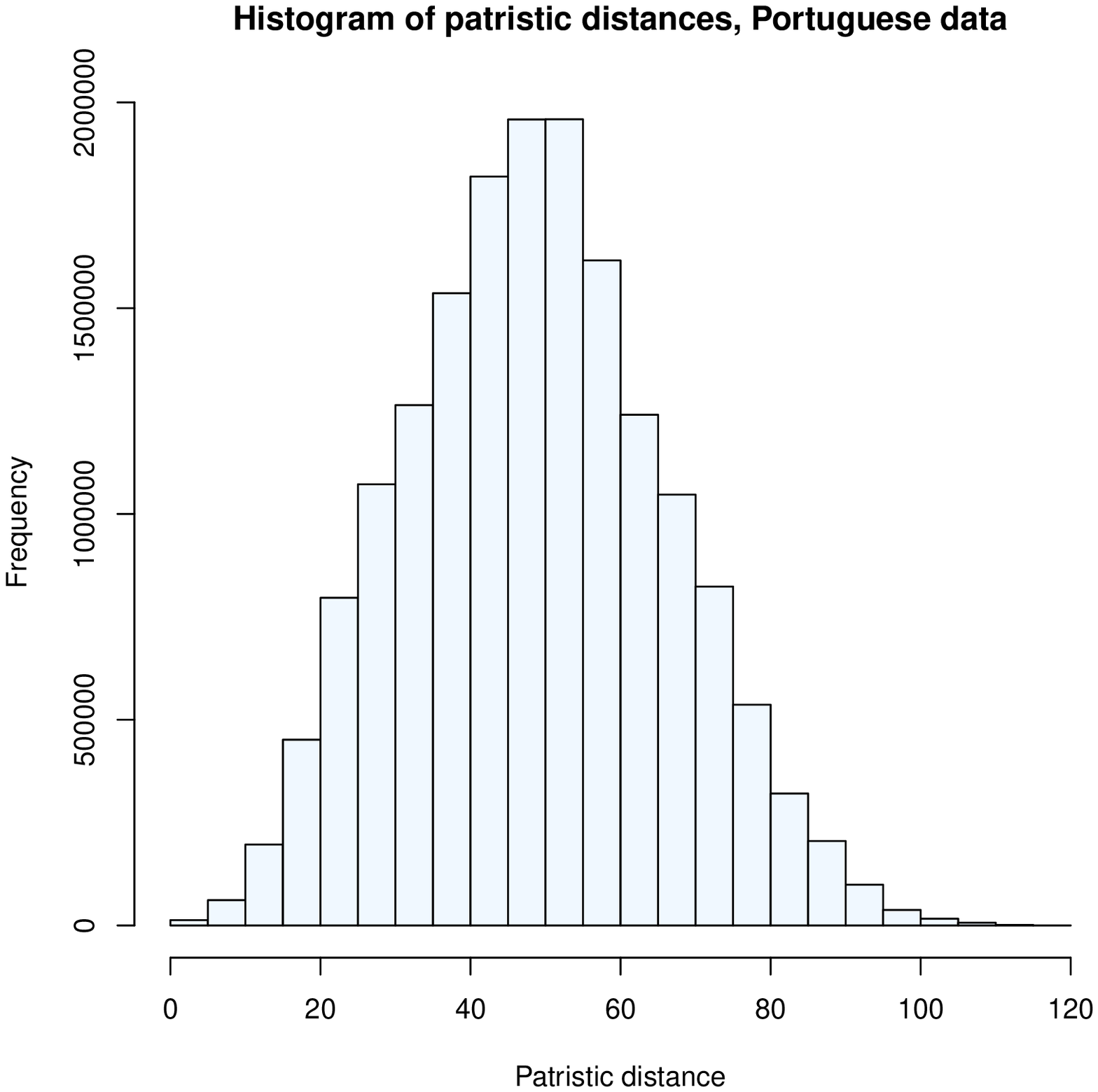}
\caption{Histogram of patristic distances in tree inferred from Portuguese dataset}
\label{fig:port_DM}
\end{subfigure}\hfill
\begin{subfigure}{.45\textwidth}
\centering
\includegraphics[width=1.2\linewidth]{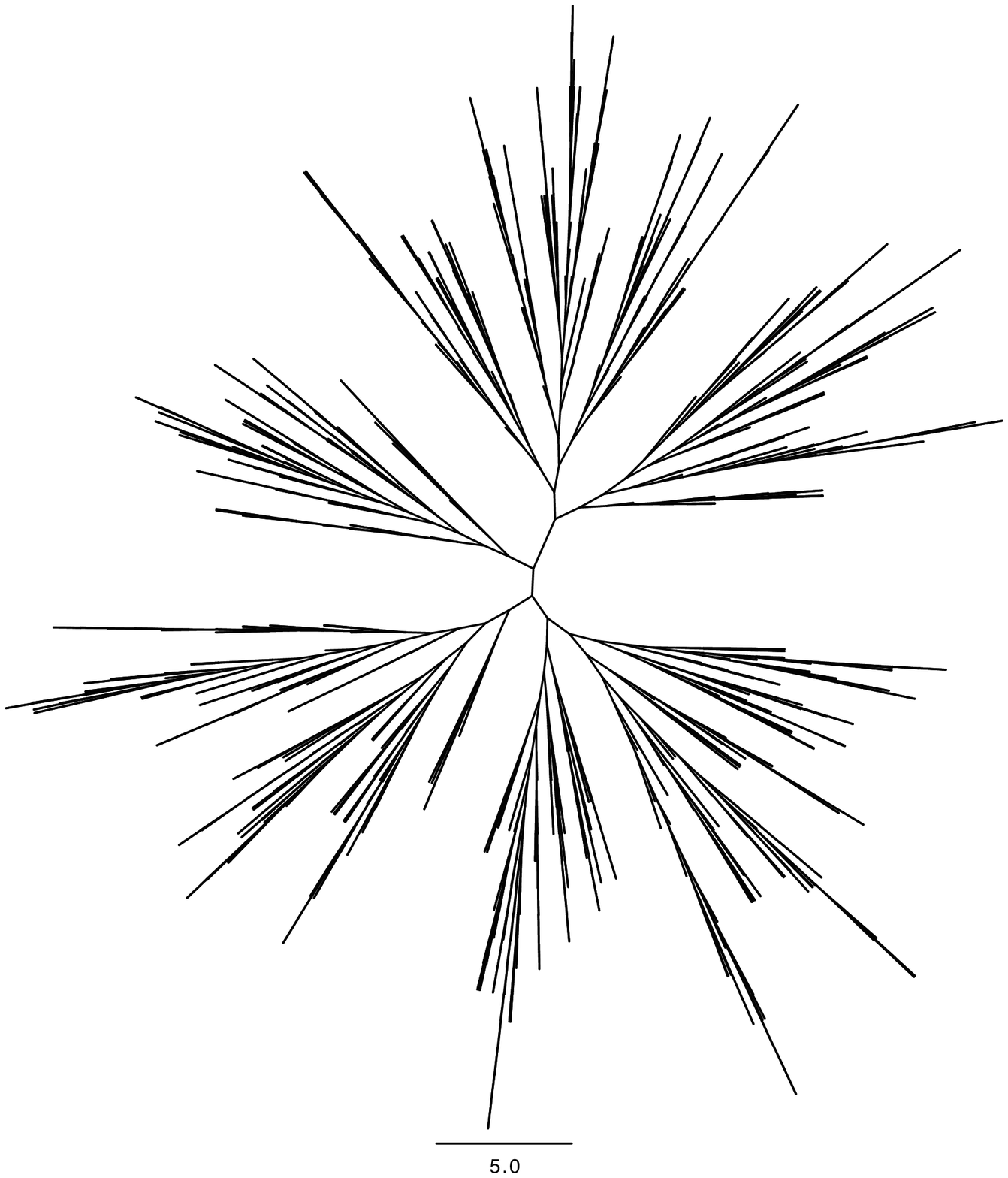}
\caption{Cladogram of tree obtained through rcolgem simulation with simulation parameters based on ABC results}
\label{fig:col_tre}
\end{subfigure}
\begin{subfigure}{.45\textwidth}
\centering
\includegraphics[width=\linewidth]{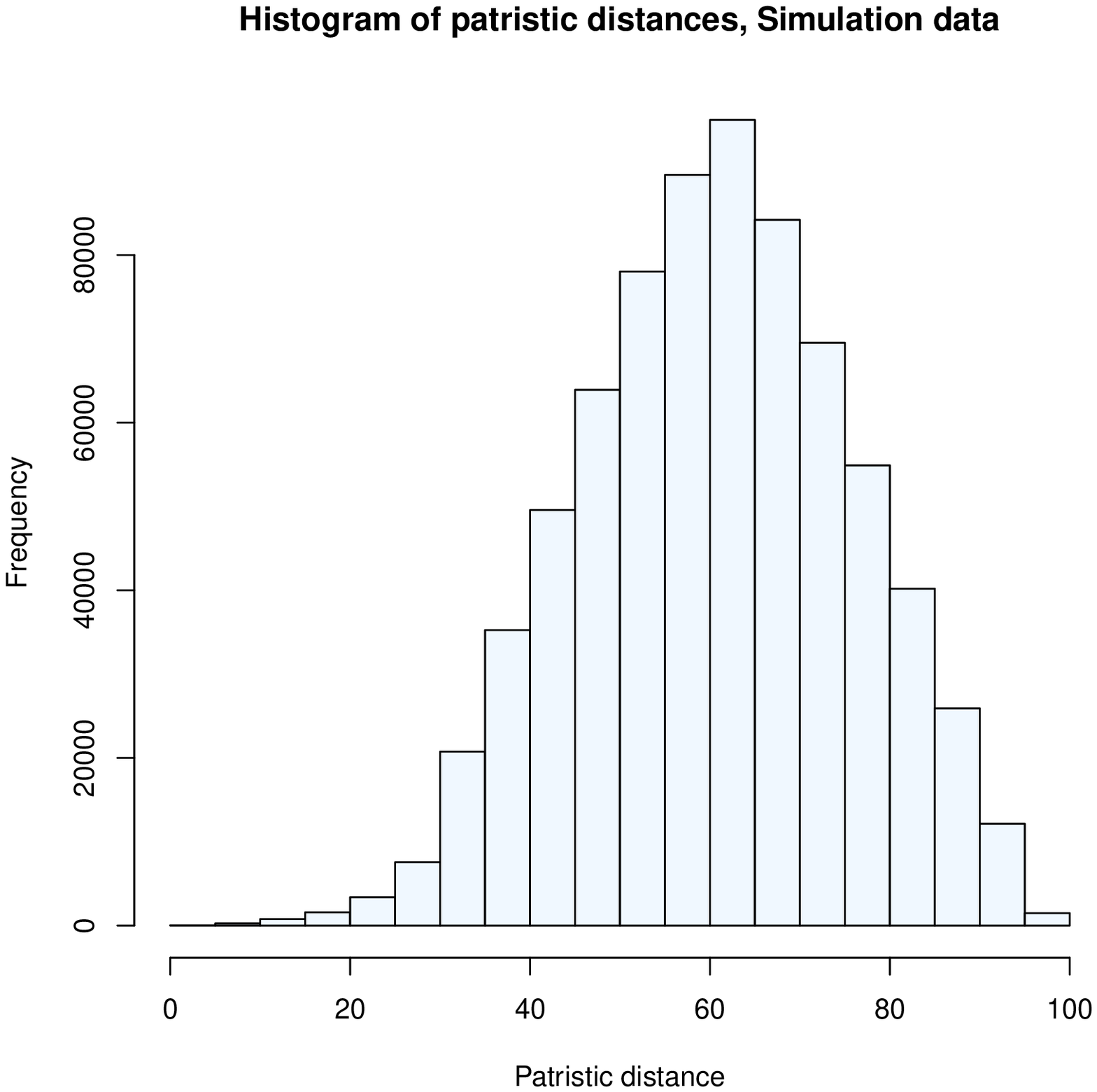}
\caption{Histogram of patristic distances in tree obtained through rcolgem simulation}
\label{fig:col_DM}
\end{subfigure}
\label{fig:tree_assess}
\begin{subfigure}{.45\textwidth}
\centering
\includegraphics[width=1.3\linewidth]{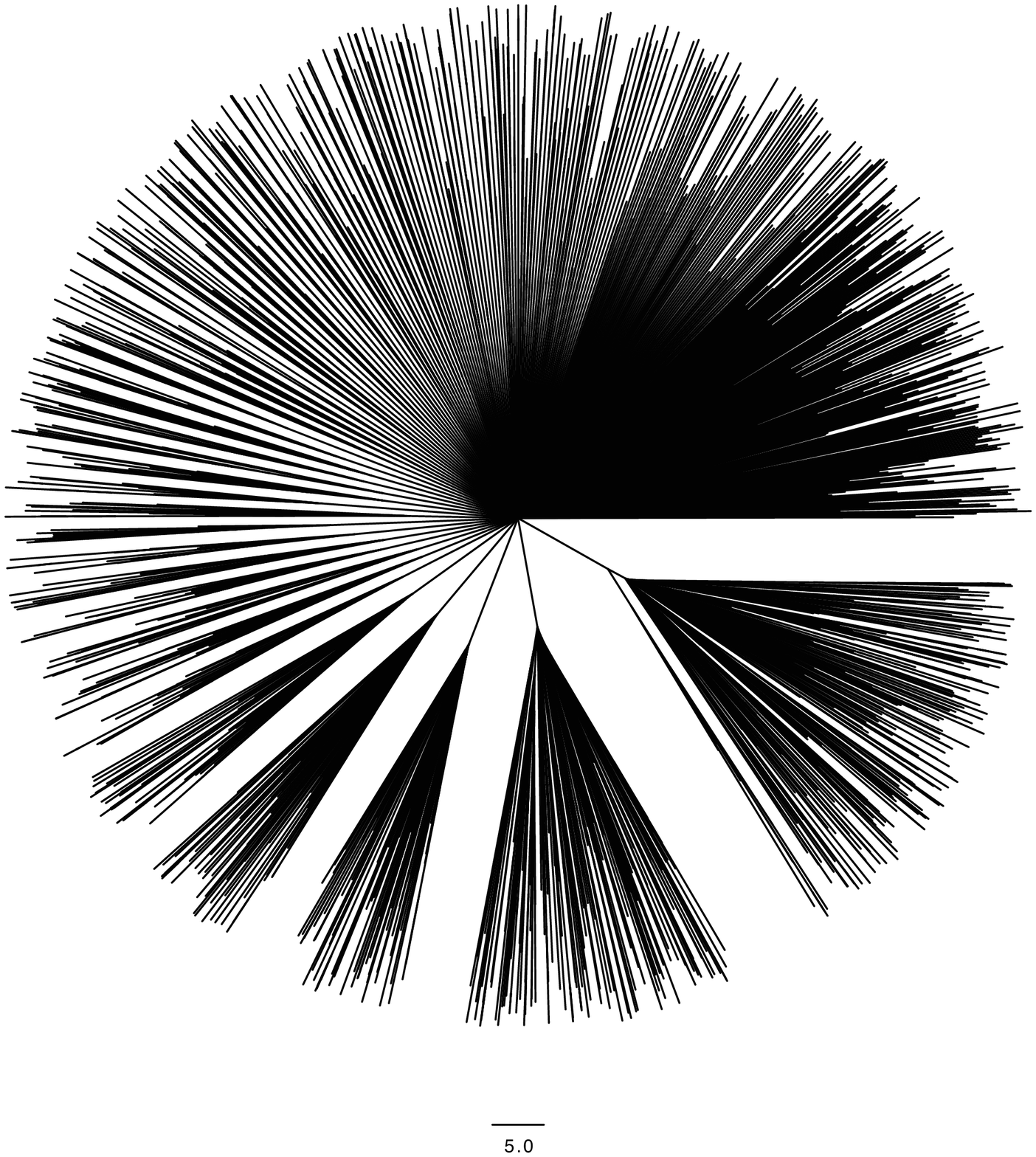}
\caption{Cladogram of tree obtained through PANGEA simulation }
\label{fig:pan_tre}
\end{subfigure}\hfill
\begin{subfigure}{.45\textwidth}
\centering
\includegraphics[width=\linewidth]{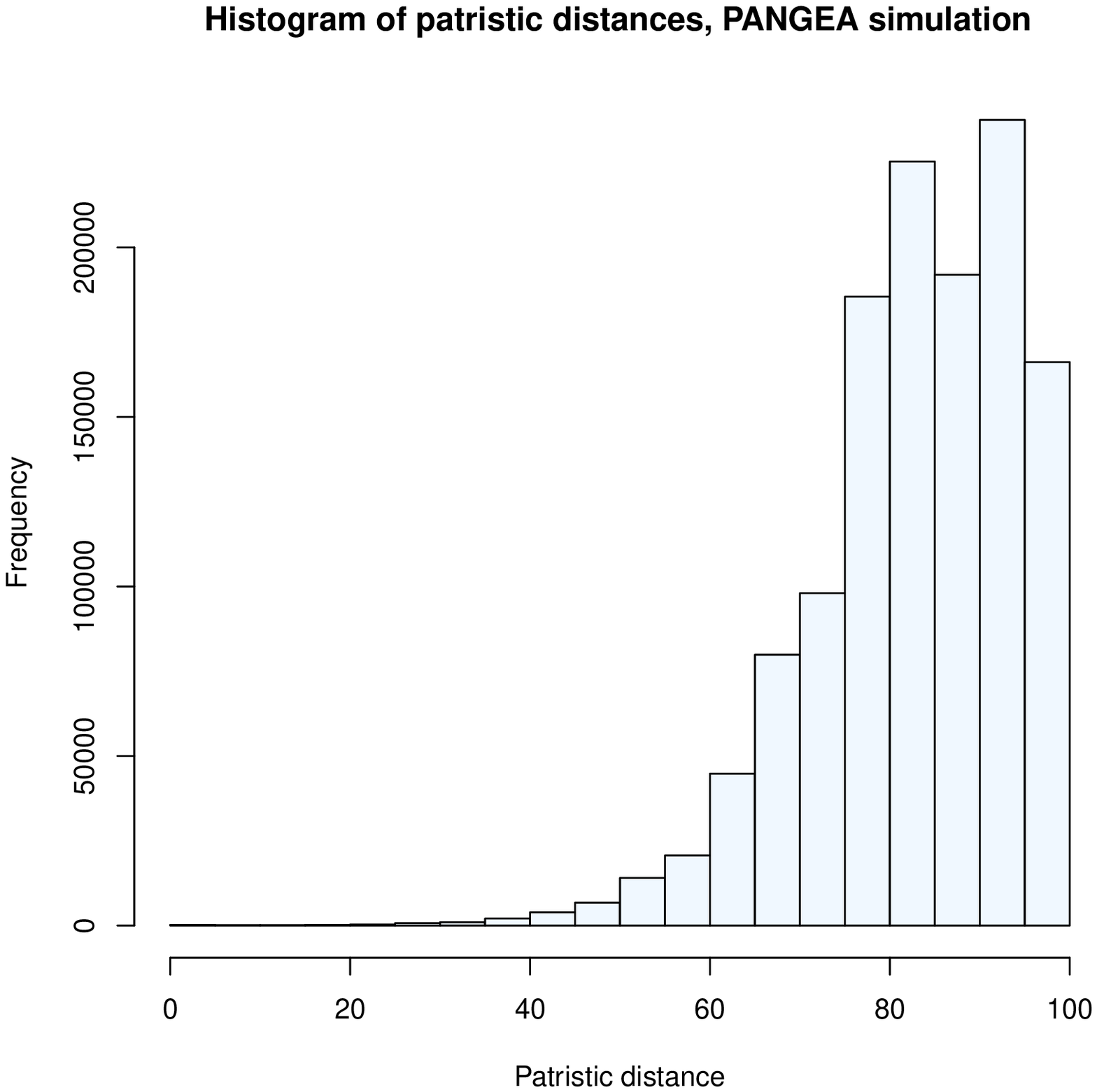}
\caption{Histogram of patristic distances in tree obtained through PANGEA simulation}
\label{fig:pan_DM}
\end{subfigure}
\caption{Comparison of topology of inferred and simulated phylogenetic trees}
\end{figure}

Figures \ref{fig:port_tre}, \ref{fig:col_tre} and \ref{fig:pan_tre} provide a visual representation of the relevant phylogenetic trees. Figure \ref{fig:port_tre} stems from the Portuguese dataset, and as such offers a baseline of desired tree topology. 
Figure \ref{fig:col_tre} presents the tree obtained through a ABC calibrated simulation, and \ref{fig:pan_tre} shows a tree from a PANGEA simulation. We demonstrate visually that the relevant tree topologies display a relatively high level of correspondence. We further investigated tree correspondence by using patristic distances \cite{phylogeotool}. The patristic distance between two leaves $l_0$ and $l_1$  is the number of changes needed to $l_0$ in order for it to become identical to $l_1$ \cite{stuessy2008patrocladistic}. Figures \ref{fig:port_DM}, \ref{fig:port_DM} and \ref{fig:pan_DM} present a histogram of the patristic distances present in the tree stemming from the Portuguese dataset the ABC calibrated simulated tree and the PANGEA tree respectively. These show that the ABC calibrated simulation tree is concordant with the tree stemming from the Portuguese epidemic.


\subsection{$T_z$-score distribution}
\label{sec:tzscore_results}
\begin{figure}
\centering
\includegraphics[width=0.6\linewidth]{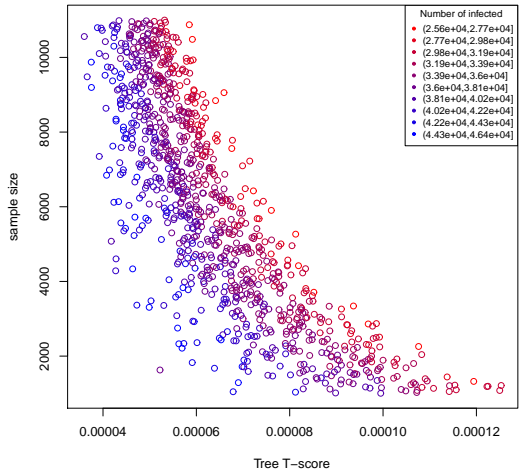}
\caption{rcolgem simulation analysis, sample size (i.e. number of leaves) against $T_z$-score, with number of infected individuals in the simulations, shown as a gradient from $2.56\times 10^4$ (red) to $4.64\times 10^4$ (blue), $N=10^5$} 
\label{fig:col_res}
\end{figure}
Figure \ref{fig:col_res} presents the result of application of our method on 1000 trees obtained from rcolgem simulations. In the figure, obtained $T_z$ scores are plotted against tree sample sizes and the number of infected individuals in the simulations, shown as a gradient from $2.56\times 10^4$ (red) to $4.64\times 10^4$ (blue). As a reminder, the number of infected individuals is determined by $\beta_i$ and $S_0$. We can clearly observe a distribution that allows distinction of the number of infected by $T_z$ scores and sample sizes. Indicating that we obtained an informative distribution through application of our method. The figure shows that a lower $T_z$ score correlates with a larger portion of the infected population not being included in the phylogenetic tree. Additionally, as sample size goes down the distribution becomes wider, indicating sufficiently large trees are necessary to allow for meaningful inferences.


\section{Discussion and Conclusions}
The presented results show that the presented $T_z$ score enables differentiation of epidemiological parameters based on phylogenetic trees. 
As such, an interesting further development would be to learn a function $f(S_0,\beta,\gamma,N)\to T^N_z$, i.e. learning the effect of epidemiological parameters on $T_z$ values in general, and from there, $f(T^N_z) \to \{\beta,\gamma\}$, i.e. constructing a function that relates obtained $T_z$ scores to possible epidemiological parameters, which we assume to be feasible, as $S_0$ should be ascertainable trough census data and epidemiological studies. \\
We would further like to extend the $T_z$ score to include a measure of uncertainty associated to inferences. A possible approach to accomplish this would be to apply our method on multiple subtrees, resulting from pruning the tree under investigation, and determining the extent to which results remain coherent with regard to the number of pruned leaves. 
Additionally,  investigating the effect of simulating epidemics using current models specifically tailored to HIV-1 (e.g. an approach where the currently prevalent CD4$^+$ models would be adapted to generate phylogenetic trees) would be an interesting further development. 
Planned further research includes adaptation of the method presented in \cite{Stacy2016} to draw coalescent probabilities from a distribution that is specific to HIV evolutionary dynamics.

We have presented a tree statistic that can be employed to assort phylogenetic trees on the basis of their underlying epidemiological parameters. By doing so, we provide a first step towards a method to infer epidemiological parameters from phylogenetic trees using coalescent theory, which would additionally be able to indicate the specific subpopulations in which diagnosis rates are low, providing a crucial tool for health policy researchers.



\section*{Acknowledgments}
Pieter Libin was supported by a PhD grant of the FWO (Fonds Wetenschappelijk Onderzoek Vlaanderen) and a grant of the VUB research council (VUB/OZR2714). 


\bibliographystyle{unsrt}
\bibliography{sigproc}

\end{document}